# Practitioners Versus Users: A Value-Sensitive Evaluation of Current Industrial Recommender System Design

ZHILONG CHEN*, Department of Electronic Engineering, Tsinghua University, China
JINGHUA PIAO*, Department of Electronic Engineering, Tsinghua University, China
XIAOCHONG LAN, Department of Electronic Engineering, Tsinghua University, China
HANCHENG CAO, Stanford University, United States
CHEN GAO, Department of Electronic Engineering, Tsinghua University, China
ZHICONG LU, City University of Hong Kong, Hong Kong
YONG LI†, Department of Electronic Engineering, Tsinghua University, China

Recommender systems are playing an increasingly important role in alleviating information overload and supporting users' various needs, e.g., consumption, socialization, and entertainment. However, limited research focuses on how values should be extensively considered in industrial deployments of recommender systems, the ignorance of which can be problematic. To fill this gap, in this paper, we adopt Value Sensitive Design to comprehensively explore how practitioners and users recognize different values of current industrial recommender systems. Based on conceptual and empirical investigations, we focus on five values: recommendation quality, privacy, transparency, fairness, and trustworthiness. We further conduct in-depth qualitative interviews with 20 users and 10 practitioners to delve into their opinions about these values. Our results reveal the existence and sources of tensions between practitioners and users in terms of value interpretation, evaluation, and practice, which provide novel implications for designing more human-centric and value-sensitive recommender systems.

**533**

*The two first authors contributed equally and are ordered alphabetically.
†Corresponding author.

Authors' addresses: Zhilong Chen, Department of Electronic Engineering, Tsinghua University, Beijing, China; Jinghua Piao, Department of Electronic Engineering, Tsinghua University, Beijing, China; Xiaochong Lan, Department of Electronic Engineering, Tsinghua University, Beijing, China; Hancheng Cao, Computer Science, Stanford University, California, United States; Chen Gao, Department of Electronic Engineering, Tsinghua University, Beijing, China; Zhicong Lu, City University of Hong Kong, Kowloon, Hong Kong; Yong Li, liyong07@tsinghua.edu.cn, Department of Electronic Engineering, Tsinghua University, Beijing, China.

## 1 INTRODUCTION

As an artificial intelligence (AI) algorithm based application specially designed for alleviating information overload, recommender systems are playing an increasingly important role in the information society [120]. The past few decades have witnessed the proliferation and increasing ubiquity of recommender systems, where recommender systems have grown to be an overwhelmingly integral part of users' online experiences and many major companies' business drivers [61]: Facebook and Twitter recommend friends, Amazon and eBay recommend products, YouTube and TikTok recommend videos, etc. Therefore, there has been an abundance of research attempts on developing recommender systems, including traditional content-based approaches [72], collaborative filtering-based approaches [101], as well as the recent deep learning enabled approaches [27, 54, 58, 120].

However, the majority of recommender system research focuses primarily on algorithmic improvements and can possibly suffer in real-world applications where user-facing performances can disconnect with highly-scored performances [51]. As a result, a comprehensive perspective that more thoroughly takes human factors into the consideration of real-world practices of recommender systems is called upon [61]. This call can be particularly important when human-AI interaction receives increasing attention in recent years in the fields of human-computer interaction (HCI) and social computing, which have long been focusing on human aspects of technology [8, 115]: with AI more and more pervasively deployed in real-world applications, humans realize the necessity of going beyond the technical aspects of AI to also attend to the complexity brought by deployment as well as social impacts [74].

Unfortunately, relatively limited research endeavors have been dedicated to examining how recommender systems, as an extremely prevalent form of AI, perform in the wild from a human-centric perspective. For the few that have, they fail to attend to the essential expectation and perception chasm in real-world deployment, which could be essential to the success of industrial recommender system design. To fill this gap, in this paper, we seek to systematically analyze how different stakeholders of recommender systems regard key values that should be supported by recommender systems in real-world practices and identify the commonalities and discrepancies with respect to each value among different stakeholders. We focus our study on two main stakeholders that are universal across different kinds of industrial recommender systems, *i.e.*, practitioners and users, to ensure the generalizability of our findings and provide implications for the design of recommender systems in general. Specifically, we ask the following research questions:

**RQ1**: What values do practitioners and users care about in terms of recommender systems? What do they think of each value respectively?

**RQ2**: What similarities and differences are shared between practitioners and users around the values of recommender systems? What tensions in values can be found?

To delve into these questions, we adopt a theoretically grounded approach, Value Sensitive Design, given its outstanding capabilities in revealing and balancing value-intense essentials across stakeholders [41]. Adopting "what a person or group of people consider important in life" [17] to define ***value***, we first leverage conceptual investigations to extract values that recommender systems should support preliminarily from prior literature. We launch two focus group discussions with expert HCI and recommender system researchers and value-sensitive frequent users of recommender systems to consolidate the values with empirical investigations. We further conduct a second round of empirical investigations of semi-structured interviews with 20 users and 10 practitioners of recommender systems to probe into their respective perspectives and behaviors towards different values and elucidate the consistency and inconsistency between the practitioner side and the user side. We take recommender systems on online-to-offline (O2O) commerce platforms as a case study to enable participants to better situate their opinions. We choose O2O commerce platforms because of their high reliance on recommender systems, their representativeness of typical recommendation scenarios, the wide adoption of their services, and our research team's access to the corresponding practitioners.

Our study highlights five key values to focus on: recommendation quality (including accuracy, diversity & novelty), privacy, transparency, fairness, and trustworthiness, and our results indicate tensions in values do exist between practitioners and users surrounding the design and practice of recommender systems. For example, we show that practitioners' and users' interpretations and emphasis of certain values can diverge, e.g., different starting points in terms of fairness. We demonstrate that although practitioners have taken various endeavors to operationalize the values, users regard them as insufficient, e.g., in terms of privacy and trustworthiness. We further analyze the sources of these tensions, *i.e.*, where these tensions originate from, and elaborate on how these tensions can be attributed to factors such as power dynamics and differences in objectives, needs, evaluations, and expectations and how they account for the gaps between practitioners' current practices of recommender systems and users' expectations. Based on these, we provide novel insights and implications for designing more human-centric recommender systems and supporting better user experiences for implementations of AI-enabled systems of this kind. The contributions of our work can be multifold:

- We reveal practitioners' and users' endeavors and concerns over key values when recommender systems are implemented in real-world scenarios through a value-sensitive approach.
- We identify practitioners' and users' tensions in values in terms of the current industrial recommender system design.
- We analyze the sources of the tensions and demonstrate that the tensions can be ascribed to factors such as diverse evaluations and distinct focuses.
- We shed light on possible solutions for mitigating the tensions and provide novel implications for the future design of recommender systems and other algorithm-driven applications.

## 2 RELATED WORK

We build our study upon two major lines of related work: 1) prior studies on values in recommender systems, the focal subject of the current study; and 2) literature on Value Sensitive Design (VSD), the theory we ground our study on. Here we refer to ***value*** as "what a person or group of people consider important in life" following Borning and Muller [17].

### 2.1 Values in Recommender Systems

Recommender systems (RSs) have long been identified as a typical human-centered information system where human factors play vital roles [61, 93]. Abundant past research has paid special attention to key values that users care about or may affect user experiences in RSs [48, 64, 66, 73]. Specifically, researchers have focused on not only pure "performance" values such as recommendation quality, but also ethical and moral values such as privacy, transparency, fairness, and trustworthiness.

**Recommendation quality.** Recommendation quality plays a foundational role in shaping users' perceptions of the functionality and utility of RSs and has received the attention of many practitioners [49, 61, 64, 91, 93, 120]. For instance, ***accuracy*** is regarded as one of the most important aspects of recommendation quality and reflects how well the algorithm can model users' interests [49, 93]. The majority of past research on RSs has endeavored to improve the accuracy of RSs [73], where approaches such as the well-known collaborative filtering method [101] and more accurate deep learning-based techniques [27, 54, 58, 120] have been proposed. To better evaluate accuracy, practitioners have developed multiple numeric indicators, e.g., Area Under Curve (AUC) and click-through rate (CTR), and often select the most appropriate indicators considering recommendation scenarios [12].

However, some researchers pointed out that the mere optimization of accuracy is insufficient to ensure recommendation quality [23, 61, 78, 86]. As such, research efforts on recommendation quality have been laid on aspects beyond accuracy. For example, ***diversity***, which represents the dissimilarity of items in a recommendation list, is gaining attention in both academia and industry [64, 86, 95, 127]. Practitioners incorporated diversity into the optimization objectives of RSs, and expected the diversified recommendations to meet users' broad-spectrum interests [122, 127]. Recently, many algorithmic techniques have been proposed to diversify recommendations, such as learning to rank [28], determinantal point process [26, 112], and post-processing [11]. Furthermore, the ***novelty*** of recommendations, referred to as the quality of the state of being novel, different, and interesting, is also growing in importance [64, 108, 114]. To achieve this, past endeavors have sought to recommend long-tailed items to users, a method which has been recognized as one of the common approaches to improve novelty [20, 108, 123].

With these continuous efforts, the quality of recommendations has been greatly improved, especially in terms of numeric metrics defined by practitioners. However, as articulated by past literature, high scores on metrics sometimes could not guarantee higher perceived recommendation quality [51, 60, 91, 113]. Thus, some researchers have made efforts to design more user-centric evaluation methods for RSs to improve the actual recommendation quality that users perceive [91, 103].

**Privacy.** Ethical and moral values such as privacy have also been highlighted in the design of RSs. For example, past research raised concerns about issues of privacy, demonstrating that users' private information can be revealed from RSs even though the systems only learn users' preferences [5, 34, 47, 62]. Thus, ample research efforts have been made to protect user privacy, addressing the potential leakage channels of the released model's parameters, recommendation results, and the overall system, and proposing multiple privacy-preserving algorithms [14, 25, 47, 79, 118].

**Transparency.** Transparency has gained increasing attention given the "black box" nature of current RSs [85, 97, 102, 106]. Past work has demonstrated that transparency, *i.e.*, the ability to make users understand how and why a recommendation is made, shapes users' decision-making processes and thus their trust in RSs [30, 87, 97, 105]. To improve transparency, researchers integrated explanations of recommendation reasons into RSs through interfaces and explainable algorithms [46, 53, 71, 81, 99, 105, 121] and underlined the importance of the quality of explanations [67].

**Fairness.** Recommender systems can also arouse issues related to fairness [4, 15, 59, 80], which can be generally divided into two categories [4, 18]: inter-user fairness, which tries to guarantee no algorithmic discrimination against a specific user or population [15, 22, 35, 125], and inter-item fairness, *i.e.*, fair exposures of different items [76, 80]. A line of works [1–3] have been dedicated to improvements towards this direction through involving multiple stakeholders into the design space for RSs and further considering their interests to improve fairness in the resulted rankings. Moreover, the importance of explaining what kind of fairness the system provides has also been highlighted [99].

**Trustworthiness.** Trust and trustworthiness are also crucial in shaping user experiences with RSs [67, 109]. Past research demonstrated that the quality of recommendations significantly influences the trustworthiness of the system [13, 90]. To enhance trustworthiness, some past literature considered incorporating social factors into RSs, for example, in the form of customer product reviews [98, 111] or by utilizing social relations in the algorithmic design [77, 88]. Other researchers endeavored to increase the trustworthiness of RSs by improving their explainability [24, 67, 119].

In short, existing literature has sought to reflect various values in recommender systems. However, it remains underexplored how these values are functioning in the wild in terms of how users' and practitioners' highlights match/mismatch. Therefore, we build on past research highlighting user aspects of RSs, for example, calls on interactive RSs [57, 63, 71] and user-centric design and evaluation of RSs [7, 36, 56, 65, 91, 99, 103], to qualitatively scrutinize tensions in values between practitioners and users of a current industrial RS, and to analyze the commonality and gaps between stakeholders with regard to the real-world deployment of RSs.

## 2.2 Value Sensitive Design

First introduced in the 1990s [40, 43], Value Sensitive Design (VSD) has been evolving over time and has developed to be an effective approach to systematically identify and elucidate stakeholders' values in the process of technology design [41]. With values defined broadly as "what a person or group of people consider important in life" [17] or more concentratedly as "what is important to people in their lives, with a focus on ethics and morality" [41], VSD discusses the relationships between technologies and human values and provides concrete theoretical, methodological, and practical groundings. To achieve this, VSD adopts a tripartite methodology integrating conceptual, empirical, and technical investigations from an interactional stance [41]. Conceptual investigations address fundamental conceptualizations informed analytically, theoretically, and philosophically, clarifying stakeholders and identifying values [41]. Empirical investigations identify how stakeholders act and apply values in practice through employing well-established qualitative and quantitative methods in social science [41]. Technical investigations examine how specific features and properties of technologies enable or hinder human activities and their values through retrospective analyses and proactive design [41]. Moreover, VSD calls for integrative and iterative usages of these three kinds of investigations to ensure robustness [41]. In this work, we follow the broader notion of "what a person or group of people consider important in life" [17] to define value. This enables us to have a more holistic view of how different kinds of values are at play in real-world scenarios and better tease out how the interweaving of usability and ethical and moral constructs shape the current industrial recommender system design.

Past research has demonstrated the effectiveness of VSD from various aspects, including identifying and legitimating stakeholders [31, 44, 83], recognizing value sources [16], enabling co-evolution of technology and social structures [45, 82], representing and eliciting key values [31, 83, 113, 116], etc. For example, Miller et al. [82] leveraged VSD to address the value tensions among diverse stakeholder groups during industrial deployment. Taking the design and development process of a corporation's groupware systems as an example, they showed how different stakeholders might hold conflicting views on similar values and how their VSD approach can alleviate value tensions among various stakeholders. Zhu et al. [124] adapted VSD to algorithmic systems and proposed Value-Sensitive Algorithm Design to address the current disconnection between research achievements by the machine learning communities and the implementation constraints and needs of the practical realities. Through bringing humans into the early stages of algorithm design, they seek to avoid biases and achieve better balances between diverse key stakeholders' stances by letting stakeholders' tacit knowledge and explicit feedback guide the direction of design iterations [124].

However, achieving such a balance is not an easy task. Views across all stakeholders can vary significantly [39], or even be contradictory [96], which sets a barrier for a universal balance and agreement. Differences in value systems, especially in cross-culture settings, can also create another layer of difficulty [6]. Moreover, stakeholders' values can change over time [75]. The intended values and the practically enacted values can also differ [9]. Sometimes even when different stakeholders' values are shared, their enactments, interpretations, and logic can also be divergent [110]. As a result, value breakdowns and assumption clashes can arise [94].

Building upon these works, we aim to examine the tensions in values in the case of recommender systems, a sub-branch of automated machine-learning algorithms where, as pointed out by prior work [124], gaps and tensions around real-world implementations are likely to take place. We adopt VSD because of its strength in surfacing issues around values [32] and manifesting and resolving tensions in values across stakeholder groups [42]. We follow VSD literature to leverage the notion of ***tensions*** in values to denote the circumstance that there could be potential differences or even "opposition" concerning values while allowing solutions that balance each side/aspect in relation to the others and enabling adjudication of the tension that holds each side's stake or aspect intact [41].

## 3 METHOD

To answer our research questions, we take recommender systems in O2O commerce platforms as a case study to assure situatedness and representativeness. We distill and consolidate the key values that are perceived to be important to be supported by recommender systems through combining literature review and focus group discussions of domain experts and users, and constantly revise them throughout our empirical investigations. We conduct semi-structured interviews to provide in-depth analysis and comparison of practitioners' and users' opinions towards these major values. Our work has been reviewed and approved by the local institutional review board (IRB).

### 3.1 O2O Commerce Platforms As a Case Study

Online-to-offline (O2O) commerce platforms [69], which provide a variety of daily sales on lifestyle services such as eating, traveling, entertaining, etc., have been growing rapidly and have achieved great success in China in recent years. For example, Meituan[1], one of Chinese largest O2O commerce platforms, acquired 510.6 million transactional users with a year-to-year growth of 13.3% and its revenue exceeded 114 billion CNY in 2020[2].

Recommender systems play an integral part in these O2O commerce platforms. Upon entering the section of a specific service, a long list of merchants/shops recommended by recommender systems is displayed. Although searching is supported, the majority of people depend on the algorithmically generated recommendation lists to choose desired merchants and shops. Through clicking the merchants or shops, users are led to pages depicting details of the merchants, where users can go through and order the items they prefer in the merchants/shops. As such, recommender systems take the role of linking users with items.

We take O2O commerce platforms as a case study for several reasons. Firstly, situating into one platform enables participants to detailedly specify their opinions and ensures consistency across interviewees. Secondly, recommender systems play an integral part in people's decision-making process in purchases on O2O commerce platforms. Thirdly, the recommendation procedures on these platforms are exemplary of the typical paradigm in many scenarios, e.g., e-commerce platforms. This representativeness enables us to delve into aspects that could suit the general usages of recommender systems, which strengthens the applicability and generalizability of our discoveries. Fourthly, the pervasiveness and wide adoption of these O2O commerce platforms make the case itself a valuable case for study. Fifthly, some members of our research teams have been actively collaborating with practitioners of O2O commerce platforms, which makes it easier for us to approach them.

[1]https://about.meituan.com/en
[2]http://media-meituan.todayir.com/20210419080003177397224\95_tc.pdf

### 3.2 Consolidations of Values in Recommender Systems

Following the Value Sensitive Design approach, we first conducted conceptual investigations of primary values that should be supported by recommender systems by reviewing prior literature as shown in Section 2.1. We defined ***value*** according to Borning and Muller's notion of "what a person or group of people consider important in life" [17]. To strengthen the generalizability of our findings and better implicate the design of recommender systems in general, we focused on two main stakeholders that are universal across different kinds of industrial recommender systems, *i.e.*, practitioners and users. Moreover, we were aware that the reference list might not be exhaustive and certain values might be missing or redundant. To prevent this, we responded to Borning and Muller's call on better contextualization and qualification of the values [17] and further conducted two focus group discussions to refine and consolidate the values with a first round of empirical investigations. Each of the focus groups consisted of 6-7 people, where we guaranteed that there were 1) at least two practitioners of recommender systems, 2) at least two HCI researchers acquainted with notable values in HCI communities who are also users of recommender systems in O2O commerce platforms, and 3) at least two frequent users of recommender systems in O2O commerce platforms who identified themselves as value-sensitive. Participants started from values derived from the conceptual investigations of a literature review to discuss in depth what values are regarded as vital for recommender systems and should thus be taken into consideration. For example, issues around whether and why each value mentioned by previous literature should or should not be underlined, whether certain values are missing and should be included, and how each value is functioning and supported were thoroughly discussed. The discussions of the focus groups lasted for 1-1.5 hours and were conducted in Mandarin. After receiving participants' oral consent, we audio-recorded the focus group discussions, transcribed and translated them to English by two native Chinese authors, and made cross-validation carefully. We retained the union of the remaining values in the two focus groups so as not to miss essential facets and constructed the primary version of consolidated values.

### 3.3 Semi-structured Interviews

To further analyze the values extracted from the focus group studies, we went deep into a second round of empirical investigations with semi-structured interviews. These interviews were conducted with both practitioners and users of recommender systems in a prevailing O2O commerce platform in April and May 2021. For participant recruitment, from the user side, we recruited 20 users of these platforms through the combination of 1) sending out posts in several Wechat groups specially for sharing for O2O commerce platforms, 2) posting WeChat Moments, and 3) putting up flyers at various places where abundant orders are completed through O2O commerce platforms to ensure the variation of the sample base. These users covered a wide range of occupations and spanned the age groups of 20-40, which constitute the major user groups of the platforms[3], among which 9 were female and 11 were male. During the interviews, we probed into the values users cared about and identified their perceptions and concerns around different aspects/values in the practice of recommender systems on these platforms, e.g., what values they care about, how they understand each value, how they think the current recommender systems support these values, how they anticipate the values should be supported, etc. For the practitioner side, we recruited 10 practitioners of recommender systems from the same pioneer O2O commerce platform. These practitioners took various roles, among which 6 were algorithm engineers, 2 were product managers, and 2 were software architects for recommender systems. 5 recruited practitioners were female and 5 were male. These interviews focused on practitioners' perceptions of and actions towards different values, e.g., what values they regard as essential to be supported, how they understand each value, how they support and evaluate the values, etc. Table 1 and Table 2 show the detailed information of the users and practitioners we interviewed, respectively. The interviews were conducted in Mandarin either through remote audio calls or in person. Each of the interviews lasted 30-75 minutes and each of our interviews with users was compensated for 50 CNY. We audio-recorded the interviews after we received interviewees' oral consent and transcribed them with the combination of automatic transcription services and manual rectifications, where sensitive and identifiable information is removed to better protect participants' privacy.

[3]https://www.questmobile.com.cn/blog/blog_125.html

Table 1. Basic information of interviewed users.

| ID | Gender | Age Group | Occupation | ID | Gender | Age Group | Occupation |
|---|---|---|---|---|---|---|---|
| U1 | F | 35-40 | Officer | U2 | F | 35-40 | Professor |
| U3 | F | 30-35 | Teacher | U4 | M | 30-35 | Designer |
| U5 | M | 30-35 | Software Engineer | U6 | M | 25-30 | Marketing |
| U7 | F | 25-30 | Algorithm Researcher | U8 | F | 25-30 | Software Engineer |
| U9 | M | 25-30 | Lawyer | U10 | F | 25-30 | Hardware Engineer |
| U11 | M | 25-30 | Government Official | U12 | M | 20-25 | Counselor |
| U13 | F | 20-25 | Algorithm Engineer | U14 | F | 20-25 | Student |
| U15 | M | 20-25 | Student | U16 | F | 20-25 | Student |
| U17 | M | 20-25 | Student | U18 | M | 20-25 | Student |
| U19 | M | 20-25 | Student | U20 | M | 20-25 | Student |

Table 2. Basic information of interviewed practitioners.

| ID | Gender | Position | ID | Gender | Position |
|---|---|---|---|---|---|
| P1 | F | Algorithm Engineer | P2 | F | Algorithm Engineer |
| P3 | M | Algorithm Engineer | P4 | M | Algorithm Engineer |
| P5 | M | Algorithm Engineer | P6 | F | Algorithm Engineer |
| P7 | F | Product Manager | P8 | F | Product Manager |
| P9 | M | Software Architect | P10 | M | Software Architect |

We first analyzed the interview transcripts utilizing open coding [29]. Two native Mandarin-speaking authors coded 10% of both practitioners' and users' interview transcripts independently and met to discuss the codes until they reached a consensus. One of them then coded the remaining transcripts and consistently met with the other to ensure agreements on the codes. A native Chinese author translated the codes and the corresponding quotes into English thereafter, where the translations were further verified by two other Mandarin-speaking authors. The whole research team then discussed thoroughly about the coded transcripts and the contents and adopted sub-categorization and constant comparison [100] to develop and refine the emerging themes. The team constantly revised and modified the elicited key values to enhance situatedness and better let practitioners and users speak for themselves [17, 41].

## 4 FINDINGS

Our semi-structured interviews enabled us to further distill key values and led us to exclude some values such as controllability and autonomy because of the limited attention reported in our conversations with participants. Eventually, we arrived at five values: recommendation quality (including accuracy, diversity, and novelty), privacy, transparency, fairness, and trustworthiness. Table 3 demonstrates the resulting consolidated values, the corresponding definitions, and representative literature on these values, respectively.

Table 3. Values in recommender systems after two rounds of consolidations.

| Value | | Definition | References |
|---|---|---|---|
| Recom-mendation Quality | Accuracy | Alignment between recommendations and user interests [93] | [27, 54, 58, 73, 93, 101, 120] |
| | Diversity | Distinction of items in a recommendation list [64] | [11, 19, 26, 28, 64, 112, 122, 123] |
| | Novelty | Distinction from previously recommended items [64] | [19, 20, 64, 108, 123] |
| Privacy | | Right that users have to keep their personal life or information non-public [62] | [5, 14, 25, 34, 47, 52, 62, 68, 79, 89, 118] |
| Transparency | | Ability to make users understand how the recommender system works [97] | [30, 53, 67, 81, 85, 87, 97, 99, 105, 107, 121] |
| Fairness | | Right to be treated equally [15] | [1–4, 15, 18, 22, 35, 59, 76, 80, 99, 125] |
| Trustworthiness | | Quality of being reliable [67, 109] | [13, 24, 67, 68, 77, 88, 90, 98, 109, 111, 119] |

With key values identified, we delve into practitioners' and users' perspectives of, opinions on, and actions towards these key values in recommender systems as reflected in our semi-structured interviews. We identify the similarities and distinctions between users and practitioners around these values correspondingly and highlight the tensions in between. Our major findings are summarized in Table 4.

Table 4. Summary of practitioners' and users' actions/opinions and tensions around different values.

| Values | Practitioner Side | User Side | Practitioner vs. User |
|---|---|---|---|
| Recommendation Quality | · Accuracy is fundamental and evaluated by numeric indicators.<br>· Diversity & novelty are complementary to accuracy and lack well-established numeric measurements.<br>· Balance between accuracy and diversity & novelty is needed. | · Accuracy is evaluated from various experience-based angles.<br>· Diversity & novelty are evaluated as a whole, based on subjective perceptions, and deficient in current practices. | · Evaluations are different: based on numeric indicators vs. subjective perceptions.<br>· Diversity & novelty should be attached more importance by practitioners. |
| Privacy | · Practitioners have taken various endeavors to protect privacy. | · Users' emphases on privacy are polarized, vary by information types, and can arise around inferences.<br>· Privacy issues could pose safety risks. | · Current privacy protection methods are not enough: sharing of hidden information, generalized anxiety, and algorithmic inference all matter. |
| Transparency | · Side effects on model performances, platform interests, and cognitive loads can be challenging for improving overall transparency.<br>· Practitioners improve transparency by enhancing explainability. | · Users are aware of the lack of transparency, but regard it as acceptable.<br>· Users need "just well" & case-dependent transparency.<br>· Current explanations are ineffective. | · Objectives and anticipations are different: general vs. just right & case-dependent.<br>· Current endeavors have not effectively made a difference. |
| Fairness | · Both inter-user & inter-item fairness are focused on.<br>· Practitioners have taken various endeavors for improvements. | · Price discrimination and inappropriate promotion are mostly concerned.<br>· Concerns can generalize across platforms. | · Starting points can be different: overall ecology vs. individual-level experiences.<br>· Gap exists between current practices & user expectations. |
| Trustworthiness | · Practitioners have taken various endeavors to cultivate trustworthiness. | · Trustworthiness can be impaired due to poor item quality, unreliable recommendation results, and the suspicious rating system. | · Liebig's Law of the Minimum leads practitioners' endeavors to maintain trustworthiness not as effective as expected. |

### 4.1 Recommendation Quality

The quality of recommendations determines to what extent the system is able to help users to distinguish their preferred item from excessive alternatives [61, 93]. In line with prior work [49, 64], we discover that three aspects of recommendation quality – accuracy, diversity, and novelty, are attached great importance by both practitioners and users. However, we also identify that discrepancies are shown regarding the interpretations, evaluations, and perceived importance of these aspects between practitioners and users.

*4.1.1 Practitioner Side.*

**Accuracy: fundamental and evaluated by numeric indicators.** Among aspects of recommendation quality, practitioners regard accuracy as the most important and underline its fundamental role in shaping the utility of recommender systems: *"Accuracy is regarded as the most important indicator ... because accuracy directly relates to commercial benefits and determines user experiences in the era of information explosion."* (P4) Practitioners always evaluate accuracy with numeric metrics based on their assumption that *"accuracy can be reflected in metrics"* (P8). Specifically, two lines of metrics are adopted: *"For offline evaluation we use Area-Under-Curve (AUC) and Group-Area-Under-Curve (GUC), and for online evaluation, we use actual business indicators such as click-through rate and purchase rate."* (P1) Specifically, for offline settings, AUC and GUC, which measure *"the ranking ability to place positive samples, which the user likes, before negative samples, which the user dislikes"* (P6), are adopted because they are regarded as *"fairer when scenarios are diverse, for example, when the numbers of items on different pages are various"* (P2) compared to other metrics such as recall and F1-score. However, these offline evaluations which rely merely on historical records and involve no new interactions with users only serve as basic thresholds for choices. It is online performances that take the final decisive role: *"The performances of offline evaluations determine whether the new model can enter the online evaluation process ... For business recommender systems, we need to generate actual business value. Model improvements have to be taken to the actual online environment for validation before releasing."* (P1)

**Diversity & novelty: complementary to accuracy and difficult to evaluate numerically.** Some practitioners also point out that the mere optimization of accuracy is far from satisfactory, and emphasize the importance of diversity and novelty in shaping recommendation quality: *"Users will leave the platform if they see homogeneous items or the recommendations can only cover parts of their needs ... Thus, we need diverse and novel recommendations."* (P4) Similar opinions are also expressed by P8: *"Accuracy comes first, but recommendations need diversity and novelty, to avoid the filter bubble phenomenon."* (P8) Compared with accuracy, defining diversity and novelty is a non-trivial task. In terms of diversity, practitioners mostly *"focus on the diversity of categories in a recommendation list"* (P7), where traditional metrics measuring statistical dispersion, e.g., entropy and number of categories, are adopted to evaluate diversity (P4, P5). As commented by P5, *"When a recommendation list consists of 10 items from four categories, it is diverse. But if all items belong to the same category, [the list] is not [diverse]."* (P5) As for novelty, it is referred to as something new and exciting to users (P2, P4, P5, P7). In this vein, novelty emphasizes the difference with what users have looked through or purchased, while diversity highlights the difference between items in a recommendation list (P5). As mentioned by P6, *"Novelty is pushing something new and exciting to the user, but it is quite difficult to infer whether a brand new item will please the user, especially when we only have historical information"* (P6). Thus, it is difficult for practitioners to find *"a plausible quantification method of novelty, especially a good numeric metric"* (P4).

**Accuracy, diversity & novelty: balance in need.** Though practitioners are aware that diversity and novelty play an essential role in *"long-term user retention"* (P4) and *"creating a good ecosystem on the platform"* (P7), how to improve diversity and novelty without impairing accuracy poses a huge challenge to practitioners. As explained by P4, *"In the short term, accuracy might be our target, which directly brings a high purchase conversion rate. But if we only chase after the current accuracy, ... users might get bored with monotonous content and do not engage long-term ... However, improving diversity and novelty might impair the current accuracy. It needs good balance."* (P4) Currently, practitioners make efforts to make better balance between accuracy and diversity & novelty (P4, P5, P6, P7, P8). For example, the diversification of recommendation lists can be accomplished by adding a rule-based adjustment to accurate sorting. As introduced by P4, *"After sorting items by their accuracy score, we will adjust the item list following some diversification rules, for example, we guarantee that there should be at least three categories in the top ten items"* (P4). However, direct improvements on novelty remain as future work due to the lack of well-established plausible metrics (P2, P5).

#### *4.1.2 User Side.*

**Accuracy: various evaluation angles.** Users regard accuracy as to what extent recommendations can match their interests (e.g., U1-U6, U9, U12, U13, U15, U17). For example, as articulated by U17, *"Accuracy is the relatedness to what I have clicked and purchased"* (U17). To evaluate accuracy, users leverage a wide range of criteria: for example, some users concentrate on the arrangement of items, e.g., *"if [the recommender system] could place an interesting item in the top 5 or 10 [positions], it is accurate; but if in the top 30 [positions], it is less accurate"* (U5). Some users evaluate accuracy from the domain of time, using characteristics such as *"time spent on finding good items"* (U2, U4, U12) as assessments. Some other users focus on users' purchase behaviors. For example, U1 uses the number of purchases as a manifestation of accuracy because *"the number of purchases reflects to what extent the recommender system can match items to users' preferences"* (U1), while users like U20 emphasize on clicks and orders: *"I click the item, and then if I place an order, it is accurate."* (U20) Others judge accuracy from the opposite side and take a more experience-based view, regarding *"the portion (I am) not interested in or would not buy ... to what extent the recommendations deviate from my interests"* (U14) as a measurement of (in)accuracy. Users' varying evaluations of accuracy often lessen the perceived importance of the items' exact rankings in the recommendation lists. As mentioned by U5, *"it does not matter if my target item moves up or down a little bit ... my preference will not change because of the ranking"* (U5). U3, U7, and U9 also share similar views, where they sometimes even pay less attention to the top ranking positions because of *"usage habits"* (U7, U9) and *"distrust of the top positions"* (U3, U5).

**Accuracy: tolerance of recommendation (in)accuracy.** As reflected by our interviewees, many users acknowledge the difficulties in providing accurate recommendations and do not make harsh requirements on recommender systems (e.g., U2, U3, U7). They highlight that as long as the recommender systems can achieve a certain level of accuracy within their limits of tolerance, it would be acceptable: *"As long as I can find something good before it runs out of my patience, it is acceptable"* (U2). This tolerance can vary from person to person: some compare more than 30 items to identify their targets (e.g., U3, U5, U7, U17), while some only glance at no than 5 items or 5 minutes (U5, U6, U14). Moreover, this tolerance is also highly context-specific and case-specific. For example, U9 reflects that *"When I am busy, I will just pick up one from the top items instantly. However, when I am free, I will hang out in the system and spend much longer time on it."* (U9)

**Diversity & novelty: identified as a whole.** In terms of diversity and novelty, users regard them as a whole, interpreting them as *"something different"* (e.g., U1, U8, U15-U18). For them, two kinds of differences are taken into consideration. Some users emphasize *"different categories"* (U15, U17), *i.e.*, the categorical coverage of recommendation lists. Some other users recognize it as *"differences from what I have known"* (U1, U8, U15, U16, U18), measuring how "out of the box" the recommendations are. Almost all participating users reflect their urgent need for these different recommendations and emphasize their irreplaceable role. For example, U7 recalls that *"I stayed at [a place] for a long* *time. I have tried most items in the list ... I really need something new."* (U7) Similar experiences are also reflected by U3, U5, U8, and U16. Moreover, some special events and cases, e.g., social events, might further enhance users' needs for diverse and novel recommendations (U5, U6, U9). As mentioned by U5, *"Once I held a party; I really want to find something new and improve the quality of the dinner."* (U5)

**Diversity & novelty: evaluated based on subjective perceptions.** To evaluate diversity and novelty, users adopt various criteria. For example, some users focus on how many different kinds of items are in a recommendation list (U15, U17), while some users appraise it by *"the portion of new items"* (U18, U19). Others break the evaluation into several small criteria and stress that diversity should be restricted to certain scopes (U16, U18). For example, as mentioned by U16, *"First of all, the system should ensure the item is of good quality. Second, the item should be new to me. Third, I actually would like to buy it"* (U16). As such, they highlight that diversity should not only ensure discrepancies but also lie within their interests. Some other users, such as U7, stress that evaluations of diversity and novelty should take a long-term perspective: *"Diversity and novelty should be assessed in a long-term manner. When I see something new, I probably would not buy it immediately. But this does not mean the recommendation takes no effect: I may just remember it and try it next time when I need changes."* (U7)

**Diversity & novelty: deficient in current practices.** Many user interviewees are disappointed to find the current recommendations are far from satisfactory in terms of diversity and novelty (e.g., U2, U10, U18). For example, some users mention that after long-term usages of recommender systems at the same place, they feel that the system is *"recommending familiar and similar items"* (e.g., U9, U11, U16), which *"makes [them] feel bored"* (U8, U9, U13) and *"limits [their] horizons"* (U3, U6). Some users attribute the boredom to personalization or possibly over-personalization (U2, U6, U7). As reflected by U2 and U6, over-personalization *"only leads me to eat what I have eaten"* (U2) and *"traps me in a boring circle"* (U6). Moreover, merely enabling the recommendation lists to cover diverse kinds of categories would not be enough to guarantee satisfaction. For example, for users who have noticed that their recommendation lists cover different categories, complaints such as *"it seems messy"* (U2, U7, U12) can arise. As explained by U15, *"In my recommendation list, some items or categories which I didn't purchase or even clicked are displayed haphazardly ... The haphazard manner means that these items show up for unknown reasons and they are arranged in an unknown order"* (U15). This seemingly messy arrangement of items does not improve users' experiences, but rather impedes their decision-making processes (U2, U7, U12, U15). Furthermore, as underscored by some user participants, diversity and novelty should be *"within the scope of interests"* (U2, U7, U12). For example, U12 complains *"I find [the system] tries to diversify the list, but it pushes too many categories that I will never order, which impairs my experience and pushes me to leave the platform"* (U12). Some users attribute the messiness to the lack of clear organizational logic and insist that diversity and novelty should be achieved in a more *"hierarchical manner"* (U6, U7, U12) that *"matches how a decision is gradually made"* (U7).

### *4.1.3 Practitioner versus User.*

**Different evaluations: numeric indicators versus user-specific, case-dependent, and perception-based assessments.** Speaking of the concrete evaluations of recommendation quality, discrepancies can be shown between practitioners and users, which can be primarily attributed to their starting points. Firstly, practitioners focus more on quantifiable numeric indicators to impartially evaluate accuracy, leveraging metrics such as AUC, GUC, CTR, and purchase rate. However, users' evaluations of accuracy are based more on their subjective perceptions, sometimes can be user-specific and stem from various angles: for example, the perceived arrangement of items, the time spent on item selection, the perceived portion of poor items, etc. Therefore, the mere use

of the current numeric metrics on records may sometimes fail to capture some aspects that may influence user experience. For example, as mentioned by U2, not only negative cases but also the specific content of the negative cases matter: *"If it pushes something that does not match my interests, I will just ignore it. But if it pushes me [food name], I will feel very sick and even eat nothing."* (U2) In cases like this, the effects of different negative cases vary greatly and highly depend on the exact user and context, which can be hard to measure through these numeric indicators. Secondly, practitioners' numeric evaluations, which focus on improving the rankings, can sometimes seem less important to users. Some users may even distrust the top recommendations altogether. For those that do trust, it barely matters if the recommendation moves up or down a bit. Therefore, the mere utilization of numeric indicators on rankings to evaluate accuracy could be questionable. Thirdly, the user-specific and case-dependent nature of users' evaluations can challenge the often "fair assumptions" of practitioners. For example, when evaluating the accuracy of recommender systems, practitioners average every user's numeric metrics and evenly treat every user. However, as has been illustrated, users' tolerance and patience in going through the recommendation lists can vary. As such, a user-adaptive and case-adaptive version of evaluations may better assess users' actual usages, the optimization of which can lead to better user experiences.

**Different relative importance of diversity & novelty.** Although both users and practitioners agree that diversity and novelty are two important aspects of recommendation quality, their perceived levels of importance compared to accuracy can vary. Specifically, users emphasize the significance of diversity and novelty in recommendations and some even place them ahead of accuracy. As reflected by U3, *"Sometimes I don't need too accurate recommendations, because I have already known my favorites. Instead, I need something I don't know or something in other categories"* (U3). In circumstances like this, users anticipate the recommender system to guide them to their potential diverse and novel favorites rather than their familiar "right" choices. However, practitioners regard diversity and novelty as optimization objectives second to accuracy and pay relatively less attention to their optimization. For example, accuracy is given the highest priority and is measured comprehensively in both online and offline settings, while diversity and novelty are not considered adequately, and there is even a lack of a well-established measurement of novelty. As such, although both practitioners and users acknowledge the importance of diversity and novelty, tensions arise from the level of importance attached. This can be largely due to the nuanced differences in practitioners' and users' expectations of recommender systems and thus discrepancies between practitioners' objectives and users' situated needs.

## 4.2 Privacy

Concerns about privacy issues of recommender systems can be mostly attributed to the characteristics of inferring users' interests based on users' personal information, e.g., user demographics and historical behavioral information [5, 34, 47]. Indeed, practitioners have taken considerable endeavors to secure recommender systems from revealing users' personal information [34, 47]. However, these endeavors sometimes fail to meet users' expectations.

### *4.2.1 Practitioner Side.*

**Endeavors to protect privacy.** Privacy is regarded as vital for recommender systems and even the overall platform (P5, P6). All of our practitioner interviewees emphasize their attention to privacy. In terms of privacy in the scenario of recommender systems, practitioners refer to users' personal information, including name, gender, telephone number, location, historical traces in the applications, etc. (P1, P4, P6-P8). To protect privacy, various actions are taken by practitioners. For example, as stated by algorithm engineers, they protect users' privacy by ensuring that all private information is collected after users accept the corresponding terms and is strictly anonymized before being fed into the recommender system (P1, P4, P6). For them, they are highly aware of privacy issues concerning data and emphasize the removal of identifiable confidential information before data processing. As reflected by P1, *"We have strict policies for data confidentiality ... Only when confidential information is removed can a table be used."* (P1) For product managers, however, they attend to not only this procedural appropriateness of removing private information, but also trying not to collect unnecessary sensitive information. As P8 states, *"what we rely on is just transactions ... there is no need for us to collect much background social information ... That [sensitive information] is beyond us."* (P8) Product managers' expertise in product operationalization helps to tease out what kinds of data are not necessary for the platforms. By avoiding the collection of unnecessary sensitive information, the possibility of breaching privacy can be further reduced from the start.

*4.2.2 User Side.*

**Polarized emphasis on privacy.** Our user interviewees display divergent and sometimes polarized emphasis on privacy. For those who are concerned with the issue, their attitudes toward sharing private information can be very conservative. For example, some users are very sensitive to the platform's data collection behaviors, and they tend to disclose as little private information as possible (e.g., U1, U5, U9, U12, U15-U18). As articulated by U2, *"if I think the information is irrelevant [to the app's current usages], I will prohibit [the app's] access to the information."* (U2) However, some other users are not so concerned with privacy issues (U4, U7, U10, U11, U14) or even have no concerns (U6, U13, U20). One of the reasons is that the platforms are characteristic of O2O commerce, which are not at high stakes from the user side and seem not so detrimental to their privacy (U4, U7, U10, U11, U14). Some users also show an understanding of recommender systems and thereby allow the system to utilize some degree of their personal information (U6, U13, U20). As reflected by U20, *"I don't mind releasing some private information ... I think this makes [the recommender system] know me better and makes [the recommender system] more convenient"* (U20).

**Privacy concerns vary in types and usages of information.** Users' concerns about privacy also relate to the specific types and usages of information. For example, most users regard income as more private because it is more personal compared to age, gender, telephone number, etc. As expressed by U2, *"income is too private and embarrassing, ... I never tell it to anyone, even my best friends"* (U2). Some users, such as U5 and U7, expect the system not to make recommendations based on income because *"I am afraid it deliberately recommends something expensive to me"* (U7). As for historical behaviors, users show higher tolerance for the use of these data. As explained by U14, *"it is acceptable that [the recommender system] uses my behavioral traces on this platform ... After all, it wants to make better recommendations for me"* (U14). However, it is emphasized by the users that the utilization should be limited to this platform and be beneficial to them. For example, some user interviewees notice that some other platforms share privacy-sensitive data among companies, which deepens their concerns about data sharing across platforms (U8, U9, U12). As commented by U8, *"I suspect that [data sharing] is a conspiracy among large companies ... [data sharing] is terrifying"* (U8).

**Privacy concerns can arise around inferences.** Some users are aware that the algorithm of recommender systems infers their personal information based on their historical behaviors (U4, U6, U7, U17). Nevertheless, attitudes towards this type of user profiling can vary. For example, some users have few concerns over user profiling and expect it to be more accurate (e.g., U3, U10, U20). As explained by U16, *"I don't care about the inference ... After all, the inference is neither real nor true."* (U16) Some take the opposite view (U2, U7, U14, U17, U19), and they feel *"insecure"* (U14) and *"beyond our control"* (U19) in terms of user profiling. Others adopt a relatively neutral attitude: they accept it as part of recommender systems but refuse to be labeled explicitly (U1, U5, U8, U9).

As explained by U9, *"If [the system] infers me, please keep the inference secret ... No one wants to be judged in public, whether by other people or by an algorithm."* (U9)

**Privacy issues could even pose safety risks.** In O2O commerce platforms, disclosing some private information, such as location and telephone numbers, is unavoidable in their nature, which could arouse users' worries about their real-world security (e.g., U2, U6, U9, U12, U15). U15 recalls a past experience: *"Once I gave a negative review to a shop, the shopkeeper called me again and again and threatened me to withdraw the review ... I was afraid that the shopkeeper would come to threaten me physically, because the shopkeeper knew my location."* (U15) Fortunately, these worries are lightened to a certain degree by the platforms' special design on privacy protection, e.g., the masked telephone number on the receipt (U3, U4, U7, U14, U19) and the use of intermediary numbers for call forwarding (U4, U7, U14).

#### *4.2.3 Practitioner versus User.*

**Permitted and cleaned data is not enough: hidden information sharing, generalized anxiety, and algorithmic inference all matter.** Both practitioners and users attach great importance to privacy and their definitions are nearly identical. However, tensions can still arise in terms of privacy issues. Practitioners emphasize that they use only permitted and cleaned data to protect users' privacy. However, users' remarks demonstrate that this is far from satisfactory. For example, users show extreme concerns about platform-to-platform information sharing. Despite noticing the issue only on certain other platforms, their suspicions can be generalized to all platforms of the kind, which may greatly affect their experiences. For example, it is likely that although a platform does well in protecting users' privacy, it is still highly suspected by users due to other platforms' violations. Furthermore, practitioners' and users' attitudes towards inferences can be contradictory. The inference of users' demographics is considered by practitioners to be part of a recommender system, which is much like assuming users' interests. Nonetheless, some users may take offense at being inferred, and even feel that it violates their privacy. As a result, despite users' recognition of practitioners' efforts, their concerns cannot be fully addressed. To a certain degree, this can be attributed to users' relatively disadvantaged positions in terms of privacy issues: they can hardly control how their data are used, whether their sensitive data are leaked, or how their demographics are inferred by algorithms. This unbalanced power dynamics in the relationships between users and practitioners then contribute to the existence of tensions on privacy issues.

### 4.3 Transparency

Despite the great success of deep learning-based recommender systems, the "black box" characteristic arouses growing concerns about how and why a specific recommendation is made, *i.e.*, the transparency of these systems [97]. Practitioners have expressed the hardships of improving transparency and tried to reduce the opacity of current systems, but the actual effectiveness users perceive can still be limited.

#### *4.3.1 Practitioner Side.*

**Challenges of improving overall transparency: potential side effects on model performances, platform interests, and cognitive loads.** Practitioners are aware of the growing calls for recommendation transparency (P1-P2, P4-P8) and recognize its importance *"throughout the overall process"* (P4) spanning inputs, system logics, outputs, etc. However, improving the transparency of recommender systems can be challenging in three ways. First, given that deep learning-based recommender systems are characterized by being accurate but opaque, practitioners are concerned that the replacement of deep models with more explainable and transparent ones will harm system performances. As illustrated by P4, *"Replacing deep black boxes with transparent models is trading accuracy for transparency, which will harm user experiences to some extent."* (P4) Second, enhancing

transparency might potentially hurt the platform's interests. For example, P6 mentions that *"if we disclose all information about how the system works, our competitors will know it and design ways to beat us"* (P6). For platforms, the black box characteristic helps protect the intellectual property of the key design in recommender systems that help them stand out among competitors, where too high a level of transparency can go against the benefits of the platform. This poses further challenges for practitioners to carefully decide the proper kinds of transparency (P6). Third, transparently illustrating how the system works could exert extra cognitive load on users, thus preventing them from enjoying smooth user experiences (P5, P6). As explained by P5, *"There are multiple stages before the overall system generates a recommendation result, including matching, ranking, re-ranking, etc. It is complicated to explain how a specific recommendation is generated ... [if we provide users with all the complicated information,] the complicated illustration can harm user experiences."* (P5)

**Explanations as an endeavor to improve transparency.** Despite multiple challenges, practitioners have been trying to improve transparency. Currently, as a first step, practitioners attempt to explain recommendation reasons to users by attaching brief explanations to each item, e.g., *"delicious and hygienic for eating"* (P1, P4). They expect that the transparency of recommendation reasons can enhance users' trust and satisfaction. As illustrated by P1, *"Explanations can help users know why the item is recommended to them ... this can make them more likely to trust the results and thus place an order."* (P1) For practitioners, the provision of explanations is believed to help users better understand the recommendation results, which enhances the transparency of the recommender systems and cultivates better user experiences.

*4.3.2 User Side.*

**Lacking but acceptable transparency.** As for users, our interviewees report a lack of transparency regarding the recommender system on the platform (U5, U7, U12, U15, U16, U19, U20). As commented by U7, *"I do not think [the system] is transparent at all ... I do not know what kind of information the system takes as the input and [the system] does not tell me its logics"* (U7). However, most interviewees are hardly disappointed by the circumstance simply because they are not even concerned about the transparency of the recommender system on the platform (U3, U6-U8, U11, U12, U16-U20). As illustrated by U12, *"Indeed, I do not care whether [the system] is transparent, because the recommendation result cannot affect my major interests. They are not high-stakes ... my core need for the recommender system here is only to get a precise list which consists of what I want."* (U12) As reflected by users like U12, in everyday scenarios like O2O commerce platforms, the recommender system performs a relatively low-stakes task on the user side. As a result, although the current system may seem to lack transparency, it seems acceptable to users.

**Calls for "just right" and case-dependent transparency.** Some interviewees, e.g., U7 and U12, point out that "just right" transparency is appreciated because excessive transparency increases their cognitive load to some degree. For example, transparency is valued only in certain specific aspects. For example, as reflected by U14, *"I have the right to know what personal information [the system] uses, for example, my locations ... but I do not care what the specific algorithm is or how it works."* (U14) From the perspective of users, the transparency of the input data is weighed more than that of the specific algorithm or model: they believe the data collected from them should be theirs, the usage of which should be informed; while the algorithm is the intelligence of the platform (U2, U6, U9, U14). Moreover, users' need for transparency can be case-dependent. As articulated by U7, *"When bad cases occur, it is the right time for [the system] to be transparent: whether it is not competent or just deliberate to make some mistakes for some profits ... this kind of transparency is important and useful"* (U7). For users like U7, transparency is not stressed or even regarded as redundant in ordinary cases. However, when dissatisfying circumstances occur, transparency is crucial for preventing users from misunderstanding or distrusting the platform.

**Ineffectiveness of recommendation explanations.** Furthermore, contrary to practitioners' hope that explaining recommendation reasons can enhance transparency and thus user experiences, many users are not even aware of those explanations or do not regard them as effective at all (e.g., U15-U20). As mentioned by U20, *"I did not notice these explanations ... and I think they are uninformative"* (U20). Similar perspectives are also reflected by U3 and U5-U8. As they point out, these explanations, e.g., *"delicious"* (U7), only claim the item's excellence, but leave out an elaboration of why it is praise-worthy or what makes it a good match for them. These explanations are thus regarded as too vague and low-quality to be persuasive enough.

#### 4.3.3 Practitioner versus User.

**Different objectives and anticipations: general vs. "just right" & case-dependent.** Although both practitioners and users attend to transparency, their objectives and anticipations differ. Specifically, practitioners ambitiously intend to improve general transparency across the whole recommendation process. Yet users do not anticipate these general improvements on transparency, but highlight "just right" and case-dependent transparency, where, in certain circumstances, too much transparency could even overwhelm them and worsen user experiences. For example, users like U12 do not intend to know the underlying mechanism because *"it is not my affair ... good results are enough in this case [of low-stakes]"* (U12). As such, practitioners' efforts can seem too general to adapt to users' emphasis on in what aspects and what circumstances transparency is regarded as important. This leads to practitioners' failure in meeting what users really call for and also explains why practitioners' first step does not take effect as expected.

**Less satisfactory first step: deficiencies of recommendation explanations.** Despite practitioners' extensive efforts to improve transparency by providing explanations for each recommended item, users can be unaware of them. For those that do, most users view these explanations as uninformative due to their vagueness and biased emphasis on positive aspects, which leads some users to progressively ignore them due to their limited help. As a result, these unhelpful explanations fail to satisfactorily improve the perceived transparency of recommender systems or bring marginal gains to user experiences.

### 4.4 Fairness

Fairness of recommender systems is gaining attention with the increase of interactions between users and the system [4, 15, 59, 80]. However, although extensive efforts have been made by practitioners, there can still be a gap between practitioners' current practices and users' expectations.

#### 4.4.1 Practitioner Side.

**Focus on both inter-user and inter-item fairness.** Practitioners regard fairness as necessary for the maintenance of a healthy ecology and thus the business: *"Unfairness can worsen the platform environment and decrease retention rates ... Fairness affects the survival of the platform."* (P8) As such, they lay focus on both inter-user fairness and inter-item fairness. As claimed by practitioners, their emphasis on inter-user fairness seeks to guarantee that there should be no difference in the quality of information accessed by users (P4, P5, P8, P9). As P9 states, *"Recommending information of varying quality to users can impair fairness. When the recommender system recommends low-quality information to a user with a poor ability on screening, it actually allocates some poor resources to him or her."* (P9) Such behavior which impairs fairness is considered unethical and can bring the platform into disrepute (P6, P9). In terms of inter-item fairness, it is identified as ensuring items of the same quality to have the same right to be displayed (P2, P4, P5, P7-P10). According to P5, *"[inter-item] fairness means whether similar items get the same amount of exposure, i.e., the number of times they are displayed and the quality of the display"* (P5). Although all kinds of practitioners acknowledge the importance of both kinds of fairness, the specific points that different kinds of practitioners attend to in their work could differ. For example, algorithm engineers stick more to quantifiable metrics, e.g., *"the amount of exposure on long-tail items"* (P2, P4, P5) and are most likely to pay special attention to inter-item fairness (P4, P5). While product managers tend to take more holistic views, focus more on the reconciliation of the relationships between the constituents of fairness and between these constituents and other factors, and highlight the notion of *"balance"* (P7, P8). As explained by P7, *"when we make strategies, we need to balance inter-item fairness, inter-user fairness, and the target of the platform ... Otherwise, [the platform ecology] cannot form a positive feedback loop, and all the efforts will be in vain"* (P7).

**Endeavors to improve fairness.** Practitioners have taken various approaches to improve the fairness of the recommender systems. In terms of inter-item fairness, many practitioners consider the long tail effect [10] as the main problem to be solved (P2, P4, P5, P8, P9). As claimed by P8, *"It is natural that a small portion of items with significantly good quality becomes the head and the rest that make up the majority become the tail."* (P8) Because of people's high preference for these head items, if merely metrics concerning accuracy are pursued, *"the recommender system may allow a small number of head items to get most of the recommended opportunities"* (P5), and most products would get few recommendations. This could lead a certain percentage of product and service providers to drop out of the platform and *"damage the ecology of the platform"* (P7, P8). Therefore, they are working consistently to ameliorate the long-tail effect, for example, providing extra support for tail items through *"giving them more opportunities for recommendations while accurately pushing them to the right people"* (P4). However, to their upset, there has been no well-established method to effectively ensure fairness among items (P2, P4, P10). In terms of inter-user fairness, practitioners' concerns lie in whether users get the same quality of information from the recommender system. As reflected by P2: *"Everyone has equal access to information, and fairness means whether everyone gets the same quality of information"* (P2).

*4.4.2 User Side.* For the user side, our user interviewees highlight that fairness affects how they perceive the platform and influences their attitudes towards the recommender system and thus the overall platform (U2, U3, U10, U14, U15, U17). For example, as articulated by U14, *"it offends me if the platform treats me differently from other users in order to make more money."* (U14) However, users can be unaware or even confused in terms of their perceptions of the current practices in supporting fairness. These perceptions can further be complicatedly intertwined with, may lead to and be led by, other values such as trustworthiness (we will discuss issues of trustworthiness in detail later in Section 4.5).

**Concerns about inter-user fairness: price discrimination.** Users' concerns about fairness also mainly focus on inter-user fairness and inter-item fairness. Regarding inter-user fairness, users' main concerns lie in price discrimination, *i.e.*, whether they would pay different prices for the same items (e.g., U3, U7, U19). For example, one frequently mentioned phenomenon is *"ripping off familiar customers"* (e.g., U2, U10, U18), where they feel that when the platform believes that a customer is accustomed to buying a specific item, it will increase its price. Such practices impair users' favor of the platform (U2, U10, U18). According to U10, *"It is becoming more and more expensive to order food on the platform. This is no good."* (U10) Although they are aware that prices are rising overall, they are skeptical of the possibilities of ripping them off because they are regular customers of the platform. Furthermore, although users show various concerns about inter-user fairness, there seems to be a lack of explanation of why and how they are treated differently in some cases (U2, U6, U12). Therefore, they have to rely more on their own "folk theories" [33, 37, 38] which further magnifies their concerns about inter-user fairness. For example, U10 mentions that *"I suspect that if someone uses an iPhone [to place an order], he/she has to pay more for the same item sometimes ... Most people believe the phenomenon exists, so do I ... I feel disappointed."* (U10)

**Concerns about inter-item fairness: inappropriate promotion.** Users are also concerned about fairness among merchants, *i.e.*, inter-item fairness (e.g., U1, U9, U17). They expect products of the same quality to have the same exposure. For example, U9 underlines that *"It would be unfair among merchants if one gets significantly more exposure for the same thing."* (U9) In this vein, users express their strong opposition against promotions by merchants' bidding (U2-U5, U8): *"I really want the recommender system to base its recommendations on the word of mouth of the items rather than the advertising fees the merchant pays for the platform."* (U2) However, the current practices are opaque, which leaves room for the possibilities of undesirable circumstances and leads the system to be unsatisfactory (e.g., U3, U4, U9). This opacity further leaves the users to depend only on their own "folk theories" [33, 37, 38] based on observations and inferences, where they widely believe that inappropriate commercial advertising exists: *"Merchants must get different recommendations, and the stores that pay more money to the platform get more recommendations"* (U9). As such, not only would fairness be hurt, but this can lend users an atmosphere of doubts and misunderstanding over the recommendation list, which worsens user experiences: *"Every time I see a new store gets referred, I always feel like they are paying a lot of money."* (U4)

**Concerns about fairness can be generalized across platforms.** Moreover, users underline that perceptions of unfairness can be generalized across platforms. Although some users may not have experienced biased cases themselves on the platform, if they are aware of possible cases in similar platforms, suspicions of unfairness can also arise (U2, U7, U9, U12). For example, regarding price discrimination, U7 mentions, *"I have observed it from other platforms, and it is common sense that platforms use big data to differentiate pricing"* (U7). Although *"I have never directly observed this phenomenon on this platform"* (U7), U7 casts doubts on it. Regarding the platform's excessive use of competitive ranking in recommendations, U2 says, *"There is no direct evidence of this platform alone"* (U2). However, U2 is convinced that *"the bidding factor accounts for an unreasonable proportion"* (U2) because *"some of the items are really good, but they wouldn't be recommended to me"* (U2).

#### *4.4.3 Practitioner versus User.*

**Different starting points: overall ecology or individual-level experiences.** Both users and practitioners recognize the importance of fairness in shaping the usages of recommender systems, yet the starting points of the two sides can be different. For practitioners, fairness is more about maintaining a healthy ecology of the platform and thus the industry. They seek to guarantee more equal and sufficient exposures among items to achieve inter-item fairness and focus on the long tail effect. For users, fairness is more related to their absolute or relative user experiences. They mind whether they are treated fairly, getting the same quality of service, and paying the same price as others: *"Users are equal to each other. If they are treated differently by the platform because of some different conditions, it is very problematic."* (U15) As such, practitioners' and users' different starting points lead to different anticipations and evaluations, which gives rise to the existence of tensions.

**Gap exists between current practices and user expectations.** Tensions can also arise in terms of the gap between practitioners' current practices and users' expectations towards the maintenance of fairness. For example, to ensure inter-item fairness, practitioners need to recommend tail products to users. However, when these recommendations fail to suit users' appetites and do not provide plausible explanations, further doubt can be cast on users' perceptions of inter-user fairness (P13). Furthermore, the personalized and customized nature of recommender systems makes it hard for direct "fair" comparisons among users: every user's recommendation lists are varying; it thus remains an obstacle to distinguish whether exposure is due to personalization or unfairness. Even worse, the lack of well-established metrics to quantify certain kinds of fairness (e.g., inter-user fairness) under personalized recommendations poses further challenges towards improving fairness (P5, P6). Moreover, users' tendency of generalizing perceptions of fairness and unfairness across platforms presents another layer of challenges. This makes it possible that although the platform has done well in ensuring fairness, if other similar platforms have not, doubts would also be raised on this platform's underlying mechanisms as well. It thus calls for improvements of the greater ecology, aggregating efforts on practitioners on multiple platforms of the same kind and possibly incorporating more stakeholders such as advisors and regulators for better regulations.

### 4.5 Trustworthiness

Regarding the significance of trust in shaping user experiences of recommender systems [50, 67, 104], practitioners have taken endeavors to improve the trustworthiness of the systems. However, users still identify several factors that undermine trustworthiness.

*4.5.1 Practitioner Side.*

**Endeavors to improve trustworthiness.** The practitioners we interview stress the importance of trust and trustworthiness in recommender systems: *"Trust determines whether users would like to use our recommender system, refer to our recommendations, and stay on the platform ... Trust is essential for the system and the overall platform."* (P6) To improve the trustworthiness of recommender systems, practitioners make various endeavors. First, they seek to ensure the reliability of the recommended shops and items. As reflected by P1, an entire team is dedicated to guaranteeing qualification, authenticity, and quality: *"One team of marketing is in charge of checking the qualification of merchants and the quality of items ... Good quality is the foundation of trust and trustworthiness."* (P1) As such, practitioners highlight the decisive role of quality in shaping trust-building for recommender systems (P1, P6). Second, they establish a trustworthiness-based rating system, where consumers can freely write reviews, as well as rating 1-5 stars for several dimensions. As mentioned by P2, *"These reviews and ratings are utilized as features of the recommender system ... We adopt the number of reviews and average ratings as criteria to filter out some items of low quality."* (P2) In this way, they assume the recommender system can automatically select trustworthy items and thereby gain more trust from users (P1, P2, P3, P7). Moreover, they notice some fraudulent behaviors by merchants, e.g., incentivizing people to make inauthentic positive reviews. Faced with these behaviors, practitioners form specialized teams filtering out fake or misleading reviews and ratings (P4, P7, P9): *"We have a special team to deal with fraudulent behaviors"* (P7). However, some practitioners point out that due to the lack of quantifiable metrics of trust and trustworthiness, it is difficult to optimize towards systems' trustworthiness and thus people's trust directly through algorithms (P5, P6).

*4.5.2 User Side.*

**Trustworthiness can be impaired due to the poor quality of items.** As reflected by users, the trustworthiness of the recommender system and the overall platform mostly results from its scale and its service infrastructure (U1, U6-U8, U10, U19, U20). However, this trustworthiness can be gradually undermined due to unsatisfactory item quality. For example, some users notice that sometimes the prices of the same items are inconsistent online and offline, which might impair their trust and prevent them from repurchasing on the platform (U6, U12, U15). As exemplified by U12, *"I know a set meal in [a restaurant] costs me 25 yuan offline, but I have to pay 35 yuan online for the set without a drink ... I think [the inconsistent prices] are deceptive ... I will never place an order there"* (U12). Some emphasize the detrimental effects of problematic recommendations, e.g., food with hygiene problems, on the trustworthiness of the recommender systems (U2, U7, U12). As mentioned by U2, *"the hygiene problems of food are very essential. Once I find the item unhygienic, I will get mad and distrustful ... I think [the platform] should be responsible for filtering the unqualified ones out."* (U2) In addition, some users recall their experiences of finding counterfeit brands, which significantly erode their trust in the systems (U6, U8, U10, U12, U15). As reflected by U6, *"Many shops claim themselves as [brand name], ... I cannot tell if they are real or counterfeit ... This makes me somewhat distrustful ... I even want to quit the platform"* (U6).

**Trustworthiness can be impaired due to unreliable recommendation results.** Moreover, some users doubt whether the recommender system is truly reflecting *their* interests (e.g., U2, U6, U12). As articulated by U12, *"I suspect how the system makes recommendations ... I doubt whether the system and the shopkeeper have some secret deals."* (U12) Some users even consider the recommender system as being manipulated, which is severely detrimental to the perceived trustworthiness of the system. As complained by U15, *"After I know that the recommendation list is not completely based on my interests or the quality, but maybe partially based on advertisers' money, I become highly distrustful of the system"* (U15). This distrust could lower down people's willingness towards further use of the system. As reflected by U2, *"The system always pushes these ridiculous items that even contradict with my interest, ... so currently I seldom depend on it"* (U2).

**Trustworthiness can be impaired due to the suspicious rating system.** The rating system also faces its share of doubt by users. Many users express a certain degree of distrust of the reviews, especially extremely positive reviews (e.g., U2-U10, U18-U20). As explained by U18, *"Overly complimentary reviews are usually fake ... Some shopkeepers will request for positive reviews using small incentives, such a bottle of water or a kebab."* (U18) Circumstances like this arouse users' suspicions on the possibility that extremely positive reviews are profit-incentivized rather than reflecting authentic feelings, which cast doubt on the overall reliability of positive reviews written by others (U2-U4, U6-U10, U12-U20). As reflected by U3, *"I suspect what portions of positive reviews are fake."* (U3) As a result, the perceived trustworthiness of the system could be lowered. Furthermore, some users judge item ratings as unreliable (U2, U4, U6, U12, U15). As articulated by U6, *"I think some ratings are inconsistent with my actual feelings ... For example, the rating of [shop name] is the highest, but it is notorious. I believe the rating is fake. Maybe the rating is adjusted by someone for some purposes."* (U6) These questionable positive reviews and ratings hinder users from making informed decisions and undermine their trust in the overall system. Instead, users depend more on photos and negative reviews, because *"they are more real and difficult to fabricate"*, which makes them more dependable (U3).

**Concerns about trustworthiness can be generalized across platforms.** Furthermore, as articulated by interviewees, impairments of trustworthiness can originate not only from the practices of the platforms' recommender systems themselves, but also from poor experiences with recommender systems on other platforms. For example, U14 expresses that doubt on recommendation results can be generalizable across platforms, *"most recommender systems contain lots of unreliable advertisements, ... so I do not believe that this recommender system is special and can truly reflect my interests."* (U14) In the same vein, other platforms' untrustworthy rating systems can also hinder users from trusting the recommender system on this platform as a result of similar generalizations. As mentioned by U12, *"[e-commerce platform name] has fake reviews and ratings, which are obviously fabricated by someone for some secret purposes ... Naturally, this leads me to think that this could happen on this platform. I thus place some doubt upon reviews and ratings on this platform."* (U12)

#### *4.5.3 Practitioner versus User.*

**Endeavors to maintain trustworthiness are not as effective as expected: Liebig's Law of the Minimum.** Despite practitioners' continuous efforts in maintaining trustworthiness, users' trust towards the recommender system has not been satisfactorily cultivated and can be vulnerable to minor deficiencies. For example, although the number of seemingly low-quality items in the recommended candidates is small, its negative influence on trustworthiness can be detrimental.

As reflected by U2, *"It pushes [item name], which is well known for its low quality. This makes me suspect whether the system breaks down or deliberately recommends it for certain reasons."* (U2) Moreover, as we have mentioned, practitioners expect the rating system to effectively collect users' perceptions of items and further examine the quality of items in a crowdsourcing way. But the real-world scenarios sometimes contradict their expectations: some shopkeepers deceive the system by incentivizing people to write positive reviews for them. Although the circumstances may not be prevalent, their existence leads users to cast doubt on the authenticity of the reviews. As a result, users' trust in the overall system can be undermined. To a certain degree, these can be attributed to practitioners' and users' different estimations, expectations, and evaluations of the influence of these factors. Users' perceptions of the trustworthiness of recommender systems relate to Liebig's Law of the Minimum [117]: it is like a Liebig's barrel, where rather than "the longest wooden bars", it is the "the shortest ones", *i.e.*, the most limiting factors and the cases they fail, determine their perceptions of the recommender system in terms of trustworthiness. While practitioners focus more on the overall or average performances, underestimate the detriment of these limiting cases on trustworthiness, and fail to increase the length of the "shortest wooden bars". Consequently, tensions occur.

## 5 DISCUSSIONS

Through looking into practitioners' and users' opinions towards various values shaping recommender systems, we have uncovered practitioners' and users' tensions in values regarding recommender systems. In this section, we reflect and highlight the key findings of our study, situate them with prior literature, and discuss possible insights and implications.

### 5.1 Identifying and Addressing Tensions in Values

Our findings suggest that tensions in values do exist when recommender systems are adopted in practice. These tensions occur not only between different values, e.g., between accuracy and diversity & novelty, transparency, and fairness, but also within a single value, e.g., in terms of the mismatches of current practices and user expectations of privacy, transparency, and fairness. These tensions also arise not only between different stakeholders, e.g., between practitioners and users, but also from the single side, e.g., within practitioners themselves.

Specifically, what we find most prominent is the tensions between practitioners and users in terms of values. Echoing Flanagan et al. [39], we discover that different stakeholders' opinions can be inconsistent or even contradictory to each other. For example, some practitioners may regard diversity and novelty as secondary to accuracy and thus attach relatively less attention to them. However, users highly stress diversity and novelty, regarding the ability to bring new knowledge through diversity and novelty as the unique characteristic that makes recommender systems special. As such, it is worth considering engaging users into the early stages of algorithm design, following prior endeavors such as Zhu et al.'s [125] in order to enable taking key human issues into account and to better mitigate tensions. Moreover, in line with Ames [9] and Robertson et al. [94], we also discover that it is possible conflicts can be shown between assumed results of technologies and real-world practices because of issues around practical implementations. For example, it is assumed by practitioners that their incorporation of explanations into recommender systems would enhance users' trust. However, users may just fail to notice these explanations, which makes it hard for the explanations to take effect. To address this, we advocate that more emphasis should be laid on how values are functioning in practice. Possible solutions to address this include adopting multi-phase collaborative design processes with stakeholders as suggested by Miller et al. [82] and attaching more focus on empirical investigations [41]. Furthermore, we also discover that even though practitioners' and users' understandings of certain values can be shared, their concrete perceptions and emphases on these values can be different. For example, in terms of accuracy, practitioners' and users' interpretations are similar. However, their evaluations can be varying, with practitioners focusing more on numeric indicators while users highlighting more diverse facets and more subjective experiences. This accords with Voida et al. [110], who underline that conflicting logic can also arise even when values are shared. To address this, it is recommended that more attention should be paid to actions rather than motivations because shared courses of action can be achieved even when stakeholders' perspectives are divergent in practice [41].

Tensions also come within a single stakeholder group. As we have shown in our findings, practitioners at different positions, e.g., algorithm engineers and product managers, can have different focuses because of their distinct job specifications and expertise. For example, in terms of fairness, algorithm engineers dedicate more to improvements on quantifiable metrics, while product managers attend more to the balance between different kinds of fairness and between fairness-related factors and other objectives. Even on the level of a single practitioner, there could be tensions between different kinds of values, e.g., between accuracy and diversity & novelty, between accuracy and transparency, and between accuracy and fairness. In these circumstances, a balance between different aspects is in need [41] and constant reflection and adjustment are called upon to cultivate better balance.

## 5.2 Human-Centric Recommender Systems

Developed to assist humans in decision making [93], recommender systems rely heavily on users' interactions, where human factors play an essential role in shaping the success of their implementation [61]. As such, researchers have a long history of taking a human-centric stance to study recommender systems, where early endeavors such as Resnick et al.'s [92] have been taken in the early 1990s. However, in recent years, relatively limited research has focused on these human-centric aspects especially when compared to the huge quantity of papers on technically improving the numeric accuracy of recommender systems. Therefore, in accordance with the historical focus and recent calls on human/user aspects of recommender systems [36, 65], we contribute a human-centric evaluation of the current industrial design of recommender systems.

Specifically, we take a value-sensitive probe into how the two main stakeholders that are universal across different kinds of industrial recommender systems, *i.e.*, practitioners and users, perceive and act towards the support of five major values in real-world practices. We discover that some points users raise align with user-centric studies of recommender systems [7, 85, 99]. However, we also extend them to uncover that users' perceived importance of values can vary significantly according to the situatedness of platform characteristics and culture: some factors prior work indicated as important (e.g., controllability [7, 56, 85]) may not be among users' major pursuits when the situated circumstances of recommender systems change; for some other factors (e.g., transparency [85, 99]), the degree that users anticipate may also differ. Therefore, we highlight the necessity to contextually understand human-centric aspects of recommender systems to support better user experiences. Moreover, building upon the few user-centric prior works that seek to elucidate users' perceptions of recommender systems [7, 56, 85, 91, 99], we advocate extending the user-centric perspective to value-sensitive stances that can effectively incorporate the opinions of other important stakeholders such as industrial practitioners. This is especially important because the real-world deployment of recommender systems could greatly suffer when there are gaps between different stakeholders' understandings and thoughts, which is exactly what we have pointed out. Therefore, we call on future research to focus on not only more human-centric design of recommender systems, but also more value-sensitive endeavors that take more stakeholders into consideration and address real-world challenges that are faced.

### 5.3 Design Implications

Through delineating practitioners' and users' tensions in values around recommender systems, we reveal vital gaps in real-world practices of recommender systems and indicate possible directions for improvements. Here we summarize major design implications inspired by our discoveries to ameliorate the current tensions and mitigate the existing gaps.

**More representative numeric metric evaluations for values.** In our findings, we demonstrate that practitioners of recommender systems rely highly on quantitative metrics to evaluate the performances of recommender systems. Some aspects, such as diversity and trust, may be hard for practitioners to improve directly because of the lack of well-established numeric metrics for evaluation. However, as we have shown, they can lay much influence on user experiences. Therefore, to enable better support of various values and thus user experiences, a prerequisite is to devise more representative numeric metrics for values that can really reflect users' situated perceptions and thoughts. This requires research attention to not only focus on improving recommendation performances on established assessment systems with more advanced frameworks, but also construct more powerful indicators to account for human-centric values so as to enable practitioners to improve on those aspects and leverage situated experiments to validate the effectiveness of these indicators. This aligns with prior research attempts such as Pu et al.'s [91] to focus on evaluations of recommender systems as well as calls by relevant communities such as the ACM Special Interest Group on Information Retrieval (SIGIR) where measurement and evaluation have been major relevant topics to contribute[4].

**More thorough user studies.** Our findings also emphasize the necessity for more thorough user studies to create better experiences. As mentioned by our practitioners, especially product managers, they indeed head for user studies to tackle certain problems (P7, P8). However, given the existence of tensions in values we identify in our findings, we deem it vital for more comprehensive user studies to be conducted. For example, we discover that some of the practitioners' efforts towards achieving the values we examine do not actually work in practice. The aspects claimed to be supported and those actually perceived could also be inconsistent, e.g., fairness. This could be due to a lack of sufficient engagement with users' field usages and opinions in real-world scenarios: some objectives for optimization are broken down by product managers with their expertise (P7, P8), and it is possible that certain aspects could be missing. Therefore, we advocate at least periodic user study-based examinations and reflections of recommender systems to identify and address the missing or newly-developed problems. This can be especially important when company policies can change with respect to time as a result of the development of the platform (P7, P8), where past endeavors towards other endeavors may not suit the new objectives.

**Longer-term investigations.** Past research on recommender systems is aware that the length of users' engagement with the system can exert difficulties on system design. However, most literature in this line focus on issues about newcomers, where the problem of cold-start has been widely discussed [21, 55, 84, 126]. We extend these studies to show that not only too limited usages but also very long term employment can be challenging (see Section 4.1). Echoing Li et al. [70] who highlight that users' usage patterns can evolve over time, we argue that practitioners of recommender systems should consider conducting longer-term investigations to satisfy users' long-term needs to retain users. We are aware that factors such as limitations on time span [70] of datasets, computing power, and fierce between-platform business competitions would increase further difficulties. However, considering that major long-term facets that deteriorate user retention may be missing, we regard longer-term investigations as necessary to ensure the sustainability of the recommender systems.

[4]https://sigir.org/sigir2021/call-for-full-papers/

**Better balance between different aspects.** As indicated in our findings, when designing recommender systems, optimizations towards different directions may come into tensions with each other to a certain degree. For example, it often requires practitioners to balance between accuracy and diversity and novelty. Enhancements in fairness could also harm accuracy and thus short-term profits. Advocates towards personalization, such as considering case-dependent and user-specific evaluations of recommendations, can contradict the notion of fairness which highlights being treated evenly and equally. Although practitioners have innate platform and company objectives as directions (P7, P8), we regard it essential to strike a better balance between these different aspects to protect users' benefits. This asks for commitments towards recognizing the aspects to be balanced and integration of aggregated dedications of practitioners at varying positions to tackle the problems, where human-centric attempts such as ours can make crucial contributions, e.g., assessing the balance from a human-centric perspective.

**More understandable illustrations of practitioners' endeavors to the public.** As we show in our findings, some users regard the current practices of recommender systems by practitioners as unsatisfactory. However, some of this can be due to a limited understanding of the concrete approaches practitioners take. Their concerns can be aroused by what they hear other similar platforms have been and depend highly on their own non-authoritative conceptions of how the system is functioning, which past literature identified as "folk theories" [33, 37, 38]. Just as has been shown by previous work, these folk theories sometimes may not be the case, where consequent tensions can arise. To deal with this, we recommend providing more understandable illustrations of practitioners' endeavors, displaying some of their high-level efforts either through publicity and media or by integrating into pushes of applications in words that are easy to understand. Although we have noticed some of such attempts[5], they could be too technical for the public to comprehend. To address this, we deem it vital for practitioners to provide illustrations of their endeavors in ways that are more comprehensible to the general public. With more knowledge on what practitioners have sought to do for guaranteeing users' stakes, users' concerns could be better alleviated and tensions between practitioners and values can be more effectively mitigated.

## 5.4 Generalizability, Limitations, and Future Work

Although our work only takes the single case of recommender systems on O2O commerce platforms in China for study, we argue that our findings (at least some of our findings) are generalizable to similar recommender systems in other scenarios. As we have illustrated, recommender systems on O2O commerce platforms align with the typical workflow of recommender systems. This has led our discoveries on tensions in values to be applicable to recommender systems in other circumstances. Moreover, not only do we focus on two main stakeholders that are universal across different kinds of industrial recommender systems, *i.e.*, practitioners and users, but the values we investigate also share a degree of prevalence among recommender systems in varied scenarios. We believe this further enhances the generalizability of many of our discoveries. However, this is not to say that we guarantee the generalizability of all our findings. Indeed, as we have discussed, some concrete features can be platform-specific and context-situated and may not apply to other platforms and other circumstances. We regard it vital to build on our work to further analyze across more platforms to enable better differentiation between platform-specific features and generalities of recommender systems.

Our study also has some limitations that are left to be tackled and uncovers several directions for future work. Firstly, our study is conducted in the Chinese context. It is likely that some discoveries may not apply to other cultures, while some other perspectives strongly held by certain cultures may fail to be reflected because of differences in belief systems. Therefore, future work could consider extending our analysis to other cultures to further distill generalities and culture-specific concerns. Secondly, future research can take more comprehensive kinds of stakeholders into account. In our study, we only consider tensions between practitioners and users of recommender systems to make it possible for our findings to generalize to other recommender systems. This can be deficient in failing to voice opinions of other stakeholders, such as merchants and shops, government officials and regulators, etc., all of whom may be valuable and informative for better situated design of recommender systems in the specific circumstance of O2O platforms. Thirdly, in our work, we observe that some values, e.g., fairness and trustworthiness, may be intertwined. Future studies can go on to detailedly examine how different values may shape or be shaped by one another. Fourthly, our work identifies gaps between conceptual investigations of literature and empirical reflections, e.g., in terms of the essence of the values of controllability and autonomy. Future work can further attend to why these gaps exist and in what circumstances these gaps would occur. Fifthly, our study also points out novel directions for improvements on specific values. For example, in terms of privacy, we identify gaps in users' understanding of privacy, which paves way for future work to consider how value conceptualization might change as designers and researchers engage with the values. Similarly, although the disclosure of numbers and addresses is inevitable for the deliveries, the collection and storage of this data may not be inevitable. This issue concerning disclosure and privacy could articulate a complex and rich design and research space for further investigations.

[5]https://mp.weixin.qq.com/s/wKZJ3toGlDQM5PKvNj7I7w

## 6 CONCLUSIONS

In this paper, we examine the key values shaping the industrial practices of recommender systems. We conduct in-depth interviews with 10 practitioners and 20 users of recommender systems to investigate their opinions and practices around recommendation quality, privacy, transparency, fairness, and trustworthiness in real-world practices. We identify tensions between practitioners and users in terms of these values, which can be attributed to diverse evaluations, distinct focuses, deficiency induced by the complexity of the real-world practices, and mismatch with user expectations. We further connect our discoveries with prior literature to discuss our contributions and provide design implications for alleviating tensions and enabling more human-centric recommender systems as well as similar AI-enabled systems.

## ACKNOWLEDGMENTS

This work was supported in part by the National Key Research and Development Program of China under grant 2020YFA0711403 and the National Natural Science Foundation of China under U1936217, 61972223, 61971267, and U20B2060.